\documentclass[fleqn,12pt]{article}
\usepackage{epsfig}

\begin{document}
\textheight 22cm
\textwidth 15cm
\noindent
{\Large \bf Mean sheared flow and parallel ion motion effects on zonal flow generation in ion-temperature-gradient mode turbulence}
\newline
\newline
J. Anderson\footnote{anderson.johan@gmail.com}, Y. Kishimoto
\newline
Department of Fundamental Energy Science
\newline
Graduate School of Energy Science, Kyoto University, Gokasho, Uji, Kyoto 611-0011
\newline
\newline
\begin{abstract}
\noindent
The present work investigates the direct interaction of sheared mean flow with zonal flows (ZF) and the effect of parallel ion motion on ZF generation in ion-temperature-gradient (ITG) background turbulence. An analytical model for the direct interaction of sheared mean flows with zonal flows is constructed. The model used for the toroidal ITG driven mode is based on the equations for ion continuity, ion temperature and parallel ion motion whereas the ZF evolution is described by the vorticity equation. The behavior of the ZF growth rate and real frequency is examined for typical tokamak parameters. It is shown that in general the zonal flow growth rate is suppressed by the presence of a sheared mean flow. In addition, with parallel ion motion effects the ZFs become more oscillatory for increasing $\eta_i (= L_n/L_{Ti})$ value.
\end{abstract}
\renewcommand{\thesection}{\Roman{section}}
\section{Introduction}
\indent
The study of plasma flows for regulating the turbulence and anomalous transport has in recent years attracted strong interest~\cite{a15}. The regulation of drift wave turbulence by sheared flows is due to shearing of the turbulent eddies and thereby reducing the spatial scales of the eddies. Of particular interest among the plasma flows are the so called zonal flows~\cite{a11} which are self generated from the background turbulence via Reynolds stress. 

Unlike the sheared mean flows which have equilibrium scale size the zonal flows are random $E\times B$ flows mainly in the poloidal direction with low frequency and thus are almost stationary compared to the time scale of the background turbulence. Since the zonal flows are quasistationary compared to the background turbulence they may keep decorrelating turbulent eddies for a relatively long time and thereby effectively suppressing the turbulent transport. 

The drift waves that are driven by gradients in the plasma density, temperature, magnetic field etc are responsible for causing turbulence and anomalous transport. Candidates of such drift waves are the Ion-Temperature-Gradient Mode (ITG) and the collisionless Trapped Electron Mode (TEM). 

In the present paper the direct effects of a sheared mean flow on the zonal flow instability are investigated analytically, however, the effects on the background turbulence itself is not considered in this work. Moreover, the effect of the parallel ion motion on the zonal flow growth rate and frequency is studied. 

The zonal flow generation from nonlinear interactions among drift waves has been extensively investigated both analytically~\cite{a51}-~\cite{a50} and in computer simulations using gyrokinetic~\cite{a22}-~\cite{a24} and advanced fluid models~\cite{a25}-~\cite{a27}. Furthermore, the effect of micro-scale electron temperature gradient (ETG) driven turbulence driven zonal flow on semi-macro scale ITG turbulence was studied by a fluid simulation~\cite{a277}. However, the direct interaction of zonal flow with a sheared mean flow and the interaction with the GAM have mostly been overlooked. There have been some previous studies on the interaction of zonal flows and mean flows using simple drift wave models~\cite{a48} and using the coherent mode coupling method~\cite{a61}. The present work extends the drift wave model to an advanced fluid model for the toroidal ITG mode turbulence while using the wave kinetic approach. The purpose of this study is to obtain a qualitative estimate of the zonal flow growth rate and real frequency, and their parametric dependence of the plasma under the influence of a sheared mean flow and parallel ion motion. The effects of sheared mean flows and parallel ion motion on zonal flows is not well known as well as how it is influencing the turbulence and in the end how this effect may change the overall transport.

However the full non-linear effects of the Geodesic Acoustic Mode (GAM)~\cite{a12} is out of the scope of the present study. The GAM is a perturbation where the electrostatic potential ($m=n=0$ mode) is coupled to the sideband density perturbations ($m=1,n=0$ mode) by toroidal effects. The GAM interacts with the zonal flow and act as a source or sink for the poloidal flow leading to an oscillatory nature of the zonal flow. The complete role of GAMs are still not well understood~\cite{a13}-~\cite{a151}. For these reasons it is of great importance to study the four component system of drift waves, sheared mean flow, GAM and zonal flow.

The methodology of the analytical model is based on the wave kinetic approach~\cite{a16},~\cite{a48}-~\cite{a50}. An algebraic equation which describes the zonal flow growth rate and real frequency including the effects of a sheared mean flow in the presence of maternal ITG turbulence including the effects of parallel ion motion is derived and solved numerically. An advanced fluid model including the ion continuity, ion temperature equations and the ion momentum equation is used for the background ITG turbulence~\cite{a28}. The generation of zonal flows is described by the vorticity equation and the time evolution of the ITG turbulence in the presence of the slowly growing zonal flow is described by a wave kinetic equation. 

It was found the generation of zonal flows was in general suppressed by a sheared mean flow. By introducing collisional damping a modest suppression of the zonal flow growth rate is found. In addition, it is found that the interaction with parallel iom momentum may reduce the ZF generation significantly.

The paper is organized as follows. In Section II the analytical model for the zonal flows generated from toroidal ITG modes including the effects of a sheared mean flow is reviewed. The analytical model is extended to include parallel ion motion in Section III. Section IV is dedicated to the results and a discussion thereof. Finally there is a summary in section V.

\section{Analytical model for interaction of mean flows and zonal flows}
In this section the model excluding parallel ion motion is introduced. The description used for toroidal ITG driven modes consists of the ion continuity and ion temperature equations. For simplicity, effects of electron trapping and finite beta are neglected in this work. Magnetic shear can, however, modify the non-linear upshift as found in Ref.~\cite{a21}-~\cite{a211} and is accordingly incorporated in the model including parallel ion motion. In this section the model of how to construct the interaction of a sheared flow and the zonal flows generated from ITG driven turbulence and the derivation of the dispersion relation for zonal flows are summarized. The method has been described in detail in Refs.~\cite{a16},~\cite{a48}-~\cite{a50} (and References therein) and only a brief summary is given here. The analysis of Ref.~\cite{a16} is closely followed and in the present work extended to be valid for an advanced fluid model. An alternative statistical approach, resulting in a modified wave kinetic equation, is presented in Ref.~\cite{a47} which also contains an extensive discussion of and comparison with the approach used here. In describing the large scale plasma flow dynamics it is assumed that there is a sufficient spectral gap between the small scale fluctuations and the large scale flow. The electrostatic potential ($\phi = e \varphi / T_e$) is represented as a sum of fluctuating and mean quantities
\begin{eqnarray}
\phi(\vec{X},\vec{x},T,t) = \Phi(\vec{X},T) + \tilde{\phi}(\vec{x},t)
\end{eqnarray}
where $\tilde{\phi}(\vec{x},t)$ is the fluctuating potential varying on the turbulent scales $x,y,t$ and $\Phi(\vec{X},T)$ is the zonal flow potential varying on the slow scale $\vec{X},T$ (the zonal flow potential is independent on $Y$). The coordinates $\left( \vec{X}, T\right)$, $\left( \vec{x},t \right)$ are the spatial and time coordinates for the mean flows and small scale fluctuations, respectively.

The wave kinetic equation see Refs.~\cite{a16},~\cite{a2} -~\cite{a8} for the generalized wave action $N_k =  \frac{4 \gamma_k^2}{\Delta_k^2 + \gamma_k^2}|\tilde{\phi}_k|^2 $ in the presence of a sheared mean plasma flow perturbing the other mean flow (the zonal flow in this case) due to the interaction between mean flow and small scale fluctuations is
\begin{eqnarray}
\frac{\partial }{\partial t} N_k(x,t) & + & \frac{\partial }{\partial k_x} \left( \omega_k + \vec{k} \cdot \vec{v}_0 \right)\frac{\partial N_k(x,t)}{\partial x} - \frac{\partial }{\partial x} \left( \vec{k} \cdot\vec{v}_0\right) \frac{\partial N_k(x,t)}{\partial k_x} \nonumber \\
& = &  \gamma_k N_k(x,t) - \Delta\omega N_k(x,t)^2
\end{eqnarray}
Here the spectral difference ($q_y << k_y$, $q_y, k_y$ is zonal flow and drift wave wave numbers respectively) is used in solving the wave kinetic equation and only the $x$ direction is considered where similar spectrum of the background turbulence and zonal flow is expected. Where $\vec{v}_0$ is the zonal flow part of the $E \times B$ velocity, and in the relation between small scale turbulence and the generalized wave action density we have
\begin{eqnarray} 
\Delta_k & = & \frac{k_y}{2}\left( 1 - \epsilon_n g + \frac{4 \tau}{3} \epsilon_n g\right) \\
\gamma_k & = & k_y \sqrt{\epsilon_n g \left(\eta_i - \eta_{i th}\right)}.
\end{eqnarray}
In the expression for the $\eta_{ith}$ the FLR effects are neglected,
\begin{eqnarray}
\eta_{i th} \approx \frac{2}{3} - \frac{1}{2 \tau} + \frac{1}{4 \tau \epsilon_n g} + \epsilon_n g\left( \frac{1}{4 \tau} + \frac{10}{9 \tau}\right).
\end{eqnarray}
Here, and in the forthcoming equations $\tau = T_i/T_e$, $\vec{v}_{\star} = \rho_s c_s \vec{y}/L_n $, $\rho_s = c_s/\Omega_{ci}$ where $c_s=\sqrt{T_e/m_i}$, $\Omega_{ci} = eB/m_i c$. We also define $L_f = - \left( d ln f / dr\right)^{-1}$, $\eta_i = L_n / L_{T_i}$, $\epsilon_n = 2 L_n / R$ where $R$ is the major radius and $\alpha_i = \tau \left( 1 + \eta_i\right)$. The perturbed variables are normalized with the additional definitions $\tilde{n} = (L_n/\rho_s) \delta n / n_0$, $\tilde{\phi} = (L_n/\rho_s e) \delta \phi /T_e$, $\tilde{T}_i = (L_n/\rho_s) \delta T_i / T_{i0}$ as the normalized ion particle density, the electrostatic potential and the ion temperature, respectively. The perpendicular length scale and time are normalized to $\rho_s$ and $L_n/c_s$, respectively. The geometrical quantities are calculated in the strong ballooning limit ($\theta = 0 $, $g\left(\theta = 0, \kappa \right) = 1/\kappa$ (Ref.~\cite{a29}) where $g\left( \theta \right)$ is defined by $\omega_D \left( \theta \right) = \omega_{\star} \epsilon_n g\left(\theta \right)$). In this analysis it is assumed that the RHS is approximately zero (stationary turbulence). The role of non-linear interactions among the ITG fluctuations (here represented by a non-linear frequency shift $\Delta\omega$) is to balance linear growth rate. In the case when $ \gamma_k N_k(x,t) - \Delta\omega N_k(x,t)^2 = 0$, the expansion of the wave kinetic equation is made under the assumption of small deviations from the equilibrium spectrum function; $N_k = N_k^0 + \tilde{N}_k$ where $\tilde{N}_k$ evolves at the zonal flow time and space scale  $\left( \Omega, q_x, q_y = 0\right)$, as
\begin{eqnarray}
\frac{\partial \tilde{N}_k}{\partial t} + iq_x v_{gx} \tilde{N}_k - k_y \left<V_E^{\prime}\right> \frac{\partial \tilde{N}_k}{\partial k_x}+ \gamma_k \tilde{N}_k = i q_x k_y \tilde{V}_E \frac{\partial N_k^0}{\partial k_x}
\end{eqnarray}
In this last expression the third term on the left hand side shows the explicit interaction of a mean shear flow $\left< V_E^{\prime} \right>$ and $\tilde{N}_k$. Here $\tilde{V}_E = iq_x \Phi$. This equation is solved for $\tilde{N}_k$ assuming that $\left< V_E^{\prime} \right>^2$ is small, by integrating by parts and obtaining an expansion in $\left< V_E^{\prime} \right>^2$ and introducing a total time derivative $D_t = \frac{\partial }{\partial t} - k_y \left< V_E^{\prime}\right> \frac{\partial}{\partial k_x}$. It is interesting to note that on the total time scale the shearing effect is explicit as $D_t k_x = -k_y \left< V_E^{\prime}\right>$. The solution can be written as
\begin{eqnarray} 
\tilde{N}_k & = & \int^{t}_{t_0} dt^{\prime} e^{-\gamma (t-t^{\prime}) - i q_x \int_{t^{\prime}}^{t^{\prime \prime}} dt^{\prime \prime}v_{gx}}i q_x k_y \tilde{V_E} \frac{\partial N_k^0}{\partial k_x} \nonumber \\
& = & - i q_x^2 k_y \Phi R_0(k_y, q_x, \Omega, \left< V_E^{\prime}\right>) \frac{\partial N_k^0}{\partial k_x}.
\end{eqnarray}

The evolution equations for the zonal flows is obtained after averaging the ion-continuity equation over the magnetic flux surface and over fast scales employing the quasineutrality and including a damping term ~\cite{a4}. The evolution equation is obtained as
\begin{eqnarray}
\frac{\partial}{\partial t} \nabla_x^2 \Phi -\mu \nabla_x^4 \Phi = \left(1 + \tau \right) \nabla_x^2 \left<\frac{\partial}{\partial x} \tilde{\phi}_k \frac{\partial}{\partial y}\tilde{\phi}_k \right> + \tau \nabla_x^2 \left<\frac{\partial}{\partial x} \tilde{\phi}_k \frac{\partial}{\partial y}\tilde{T}_{ik} \right>
\end{eqnarray}
where it is assumed that only the small scale self interactions are the important interactions in the RHS~\cite{a31}. Using typical tokamak parameters ($T_i = T_e = 10 kev$, $n_i = n_e = 10^{20}m^{-3}$, $r = 1m$, $R = 3m$) $\mu = 0.78 \nu_{ii} \sqrt(r/R)$ and $\nu_{ii} = 10^{-12} n_i/T_i^{3/2}$ and $\nu_{ii}$ is the ion-ion collision frequency, $T_i$ is the ion temperature in electron volts. Using typical tokamak parameters it is found that $\mu \approx 50$.
Expressing the Reynolds stress terms in Eq. 8 in $N_k$ we obtain
\begin{eqnarray}
\left(-i \Omega - \mu q_x^2 \right) \Phi = \left(1 + \tau  + \tau \delta \right) \int d^2 k k_x k_y |\tilde{\phi}_k|^2
\end{eqnarray}
where $\delta$ is a $k$ independent factor
\begin{eqnarray}
\delta = \frac{\Delta_k k_y}{\Delta_k^2 + \gamma_k^2} \left(\eta_i - \frac{2}{3} \left(1 + \tau \right)  \epsilon_n g\right).
\end{eqnarray}
Utilize equations 7, 9 and 10 gives,
\begin{eqnarray}
\left(-i \Omega - \mu q_x^2 \right) = - q_x^2 \left(1 + \tau  + \tau \delta \right) \frac{\Delta_k^2 + \gamma_k^2}{4 \gamma_k^2} \int d^2 k k_y^2 k_x \frac{\partial N_k^0}{\partial k_x} R_0.
\end{eqnarray}
Here the response function ($R_0$) is, considering only the first two even terms in the expansion
\begin{eqnarray}
\bar{R}_0 & = & \frac{1}{\gamma_k - i(\Omega-v_{gx} q_x)} \\
R_0 & = & \bar{R}_0 + \bar{R}_0^2 D_t \bar{R}_0 D_t.  
\end{eqnarray}
In contrast with Ref.~\cite{a49} it is not assumed that the short scale turbulence is close to marginal state (or stationary state, $\gamma_k$ is small). Integrating by parts in $k_x$ and assuming a monochromatic wave packet $N_k^0 = N_0 \delta\left(k - k_0\right)$ gives
\begin{eqnarray}
\left(\Omega + i \mu q_x^2 \right) = -  i q_x^2 \left(1 + \tau  + \tau \delta \right) \frac{\Delta_k^2 + \gamma_k^2}{4 \gamma_k^2} k_y^2 R_0(k_y, q_x, \Omega, \left< V_E^{\prime} \right> ) N_0 
\end{eqnarray}
Here, $\Omega$ is the zonal flow growth rate and real frequency, $q_x$ is the zonal flow wave number and $k_y$ is the wave number for the ITG mode. The real part of the sheared mean flow dependent term can now be written in the $k_{\perp} << 1$ limit as found in Ref.~\cite{a48},
\begin{eqnarray} 
Re(R_0) & \approx & \frac{\gamma_k}{\gamma_k^2+(\Omega-q_xv_{gx})^2} \nonumber \\
& - & 12 q_x^2 (q_f^2 \Phi_f)^2 k_y^2\left( \frac{\gamma_k}{\gamma_k^2+(\Omega - q_x v_{gx})^2}\right)^5.
\end{eqnarray}
This result is valid in the weak shear limit $\gamma_{ZF} > \left< V_E^{\prime}\right>$. In later numerical calculations the full expression is retained. In expressing the zonal flow growth in dimensional form making use of the relation $\left(\Delta^2_k + \gamma_k^2\right)/\left( 4 \gamma_k^2 \right) N_0 = |\tilde{\phi}|^2$ it is assumed that the mode coupling saturation level is reached~\cite{a40} 
\begin{eqnarray}
\tilde{\phi} = \frac{\gamma}{\omega_{\star}}\frac{1}{k_y L_n}
\end{eqnarray}
When calculating the group velocities the FLR effects in linear ITG mode physics is important and given by the real frequency and growth rate as follows from Ref ~\cite{a49}
\begin{eqnarray}
\omega_r & = & \frac{k_y}{2\left( 1 + k_{\perp}^2\right)} \left( 1 - \left(1 + \frac{10\tau}{3} \right) \epsilon_n g - k_{\perp}^2 \left( \alpha_i + \frac{5}{3} \tau \epsilon_n g \right)\right)  \\
\gamma & = & \frac{k_y}{1 + k_{\perp}^2} \sqrt{\tau \epsilon_n g\left( \eta_i - \eta_{i th}\right)}
\end{eqnarray}
where $\omega = \omega_r + i \gamma$. The group velocities ($v_{gj} = \partial \omega_r/\partial k_j$) are in the long wavelength limit ($k^2_{\perp} << 1$) given by,
\begin{eqnarray}
v_{gx} & = & - k_x k_y \left(1 + \left( 1 + \eta_i\right) \tau - \left(1 + \frac{5 \tau}{3} \right) \epsilon_n g \right) \\
v_{gy} & = & \frac{1}{2} \left( 1 - \left( 1 + \frac{10 \tau}{3}\right) \epsilon_n g\right).
\end{eqnarray}
\section{Analytical model including parallel ion motion}
It is known that parallel ion motion effects on the background turbulence growth rate is only slightly modified, whereas, there is often a significant effect on the real frequency (significant increase in $|\omega_r|$). The zonal flow dispersion equation is explicitly dependent of the background real frequency, group velocity and growth rate and it is now expected that there is significant effect on the zonal flow generation dependent on the parallel ion motion. Accordingly, the previous model for the generation of zonal flows from ITG background turbulence is extended to include the equation of motion for the ions. The model for the drift waves consists of the following equations:
\newline
ion continuity equation
\begin{eqnarray}
\frac{\partial \tilde{n}}{\partial t} - \left(\frac{\partial}{\partial t} - \alpha_i \frac{\partial}{\partial y}\right)\nabla^2_{\perp} \tilde{\phi} + \frac{\partial \tilde{\phi}}{\partial y} - \epsilon_n g \frac{\partial}{\partial y} \left(\tilde{\phi} + \tau \left(\tilde{n} + \tilde{T}_i \right) \right) + \frac{\partial \tilde{v}_{i||}}{\partial z} = \nonumber \\
- \left[\phi,n \right] + \left[\phi, \nabla^2_{\perp} \phi \right] + \tau \left[\phi, \nabla^2_{\perp} \left( n + T_i\right) \right]
\end{eqnarray}
ion energy equation
\begin{eqnarray}
\frac{\partial \tilde{T}_i}{\partial t} - \frac{5}{3} \tau \epsilon_n g \frac{\partial \tilde{T}_i}{\partial y} + \left( \eta_i - \frac{2}{3}\right)\frac{\partial \tilde{\phi}}{\partial y} - \frac{2}{3} \frac{\partial \tilde{n}}{\partial t} = - \left[\phi,T_i \right] + \frac{2}{3} \left[\phi,n \right]
\end{eqnarray}
parallel ion momentum equation
\begin{eqnarray}
\frac{\partial \tilde{v}_{i||}}{\partial t} & = & - \left(\frac{\partial \phi}{\partial z} + \tau \frac{\partial}{\partial z}\left( \tilde{n} + \tilde{T}_i\right)\right) - \left[\phi,\tilde{v}_{i||} \right].
\end{eqnarray}
Here $\left[ A ,B \right] = \partial A/\partial x \partial B/\partial y - \partial A/\partial y \partial B/\partial x$ is the Poisson bracket. The quantities are normalized in the same fashion as above with $\tilde{v}_{i||} = (L_n/\rho_s)  v_{i||}/c_s$. The electrons are assumed to be Boltzmann distributed. Note that, for the zonal flows $k_{\parallel} = 0$ and Eq. 23 is identically zero and Boltzmann distributed electrons cannot be used, instead the same model as earlier is employed c.f Eq. 8. The dispersion relation for the ITG mode resulting from Eqs 21 - 23 is then
\begin{eqnarray}
\left[1 + k_{\perp}^2 \left(1 + \frac{5 \tau}{3} \right) \right] \omega^2 - \left[ 1 - \epsilon_n \left( 1 + \frac{5 \tau}{3} + \alpha_r\right) - k_{\perp}^2 \tau \Gamma \right. \nonumber \\
\left. -i \left( \frac{\epsilon_n s}{2 q} \left(1 + \frac{5 \tau}{3} \right) \right) \right]\omega k_y \nonumber \\
+ \left[ \epsilon_n \left( \Gamma - \alpha_r + \frac{5 \tau^2}{3}k_{\perp}^2 \left( 1 + \eta_i\right)\right) + i\left( \frac{\epsilon_n s}{2 q} \Gamma \right) \right] k_y^2 = 0
\end{eqnarray}
Here $\alpha_r = \frac{5 \tau}{3}$, $\tau = T_e / T_i$ and
\begin{eqnarray}
\Gamma = \tau \left( \eta_i - \frac{2}{3}\right) + \frac{5 \tau}{3} \epsilon_n \left( 1 + \tau \right)
\end{eqnarray}
Here, the solution have been found using an approximate eigen-mode function in the form of the lowest order Hermite polynomial ($n=0$).
\begin{eqnarray}
\delta \phi \propto e^{-z^2/2\sigma^2}
\end{eqnarray}
where 
\begin{eqnarray}
\sigma = \frac{i \epsilon_n}{k_{\perp} |s| q \omega}
\end{eqnarray}
Here, $Re(\sigma^{-2}) > 0$ has been imposed as a causality requirement. Now, we proceed as in Ref.~\cite{a49} and obtains a modified 3rd order dispersion relation for the zonal flow growth rate and real frequency.
\begin{eqnarray}
\left(\Omega + i \mu q_x^2 \right)\left(\Omega - q_x \frac{\partial \omega_r}{\partial k_x}\right)^2 = - q_x^2 \left(1 + \tau  + \tau \delta \right) k_y^2 R_0 |\tilde{\phi}_k|^2 \Omega 
\end{eqnarray}
where $\omega_r$, $\gamma_k$ and the numerical derivative $\frac{\partial \omega_r}{\partial k_x}$ are found from the dispersion relation (Eq. 24), $R_0$ is the same factor as before and where the factor $\delta$ now is modified to,
\begin{eqnarray}
\delta = k_y \frac{\omega_r - \frac{7}{3}\tau \epsilon_n g k_y}{(\omega_r - \frac{7}{3}\tau \epsilon_n g k_y)^2 + \gamma_k^2} \left(\eta_i - \frac{2}{3} \left(1 + \tau \right)  \epsilon_n g\right).
\end{eqnarray}
In present model extended for including parallel momentum and mode structure there are some limitations, it is only a change in the linear drift wave physics and the possible non-linear coupling term in Eq. 28 is not included and analogously the generalized action density derived in~\cite{a49} for the background turbulence is only approximately correct in the long wavelength limit.
\section{Results and discussion}
The model in Ref.~\cite{a49} is expanded by including the effects of interaction with mean flows and the effects of parallel ion motion for the zonal flow generation. An analytical dispersion relation is derived according to the wave kinetic equation approach that takes into account the effect of a shear flow interacting with the generation of zonal flows. The plasma parameter dependence on the zonal flow growth rate and real frequency is explored for comparison purposes with the explicit dependence of mean flow shear ($\left< V_E^{\prime}\right> \propto q_f^2 \Phi_f$) given special attention. Here $q_f$ is the wavenumber and $\Phi_f$ is the electrostatic potential of the sheared mean flow. First the results from the analytical model derived from the wave kinetic approach is presented including the effect of a sheared mean flow (solution to the dispersion relation Eq. 12).

In Figure 1 the zonal flow growth rate and frequency as a function of $\eta_i$ with the collisional damping as a parameter is displayed. The other parameters are $q_x = 0.3 = k_x = k_y$, $\tau = 1$ and $\epsilon_n = 1$ with $q_f = 0$ and $|\Phi_f| = 0$. The results are shown for $\mu = 0$ (plus), $\mu = 10$ (boxes) and $\mu = 50$ (rings). In the figure the growth rates are positive while the real frequencies are negative for the zonal flow. In the case of no damping $\mu = 0$, almost no explicit $\eta_i$ dependency on the zonal flow growth rate and real frequency appears, whereas, for increasing damping ($\mu$) modest effects on the growth rate and real frequency of the zonal flow $\Omega$ are exhibited. 

\begin{figure}
  \includegraphics[height=.3\textheight]{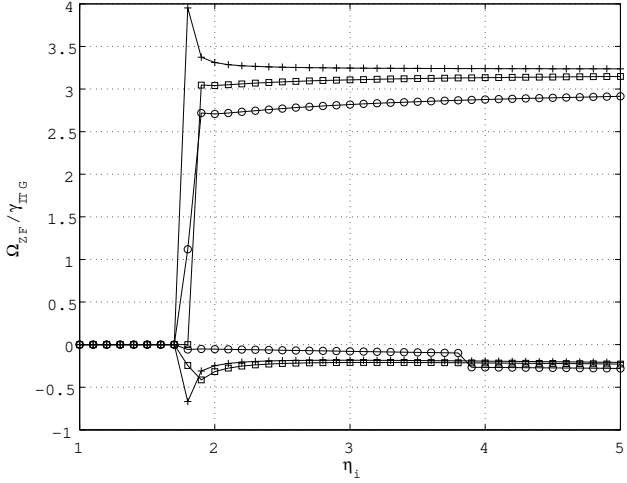}
  \caption{the zonal flow growth rate and frequency  as a function of $\eta_i$ with collisional damping ($\mu$) as a parameter. The other parameters are as $q_x = 0.3 = k_x = k_y$, $\epsilon_n = 1.0$ and $\tau = 1$, $q_f = 0$ and $ |\Phi_f| = 0$. The results are shown for $\mu = 0$ (plus), $\mu = 10$ (boxes) and $\mu = 50$ (rings).}
\end{figure}

Next in Figure 2 and 3, the zonal flow growth rate (Figure 2) and real frequency (Figure 3) as a function of $\eta_i$ with $|\Phi_f|$ as a parameter is shown. The other parameters are as in Figure 1 with $q_f = 0.1$. The results are shown for $|\Phi_f| = 0$ (plus), $|\Phi_f| = 100$ (boxes) and $|\Phi_f| = 500$ (asterisk). The effect of a mean flow on zonal flow generation is qualitatively similar to that of damping, however, in the mean flow case the growth rates and real frequencies approach the case with no flow $|\Phi_f| = 0$ for large values of $\eta_i$. In addition, unlike the damping, the sheared mean flow is also responsible for a significant suppression of the zonal flow growth rate for small $\eta_i$-values to a certain minimum level determined by the background turbulence. If the background is considered to be close to the marginal state a total suppression of the zonal flow growth rate is found. Note that the effect of mean flow is included in the product of $q_f^2$ and $|\Phi_f|$, namely the shearing rate. The resulting zonal flow growth rate reflects that the solution to the dispersion relation consists of two branches where one branch is not depending on the mean flow damping and another branch that is significantly damped close to marginal stability. The behavior of the zonal flow growth rate is indicated in Eq. 15 where the mean flow damping is explicitly found and that it is reduced with increasing $\eta_i$-values (i.e. increasing ITG growth rate). If the sheared mean flow is allowed to interact with the background and act as $E\times B$ shear the linear ITG growth rate will be reduced resulting in a reduction of the zonal flow growth rate. Moreover, in Eq. 15 it is indicated that a reduction in the linear growth rate will also result in a stronger direct effect of a sheared mean flow damping on the zonal flows.

\begin{figure}
  \includegraphics[height=.3\textheight]{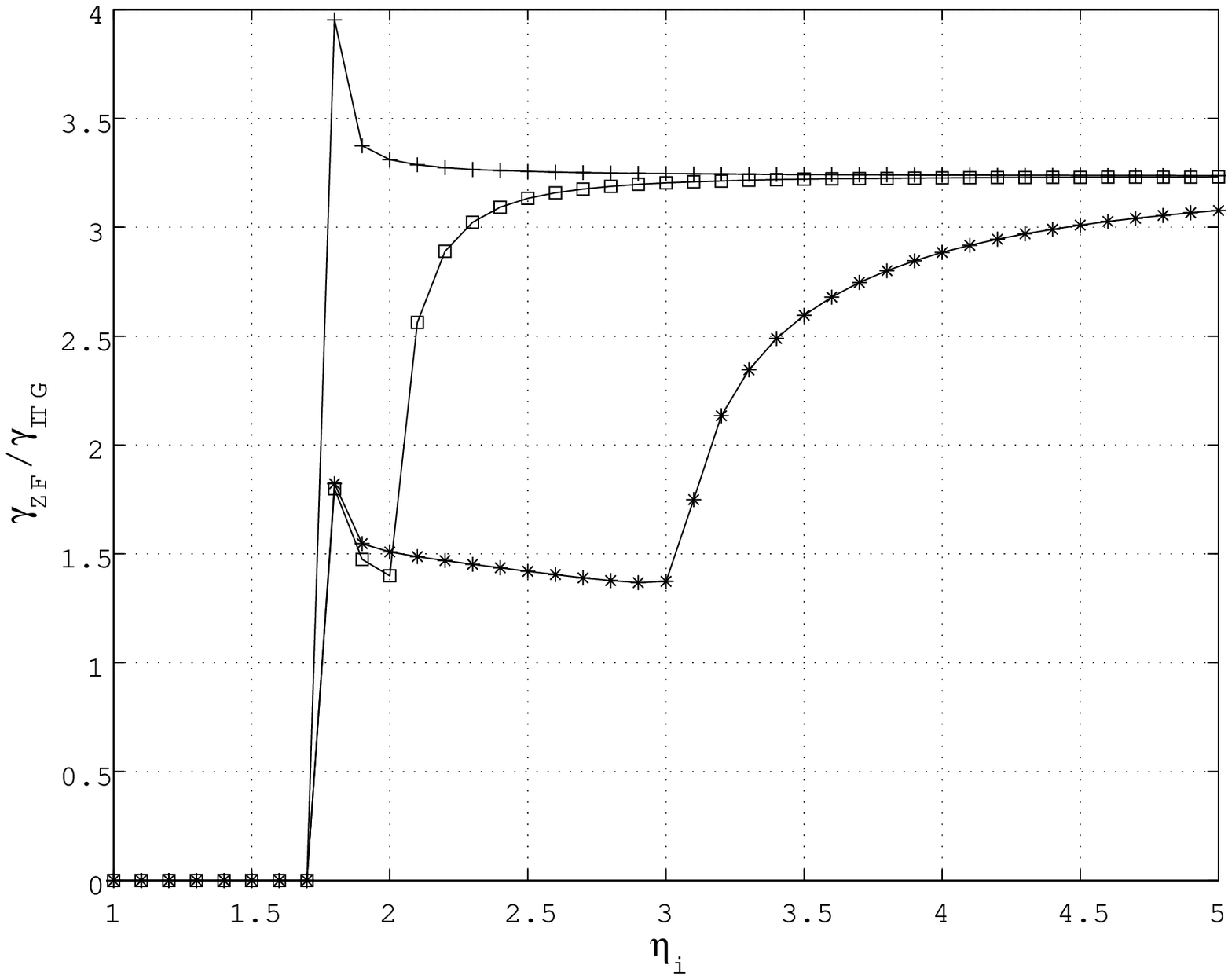}
  \caption{(2 and 3) The zonal flow growth rate (Figure 2) and frequency (Figure 3) as a function of $\eta_i$ with $q_f |\Phi_f|$ as a parameter. The other parameters are as $q_x = 0.3 = k_x = k_y$, $\epsilon_n = 1.0$ and $\tau = 1$. The results are shown for $q_f |\Phi_f| = 0$ (plus), $q_f |\Phi_f| = 100$ (boxes) and $q_f |\Phi_f| = 500$ (asterisk).}
\end{figure}

\begin{figure}
  \includegraphics[height=.3\textheight]{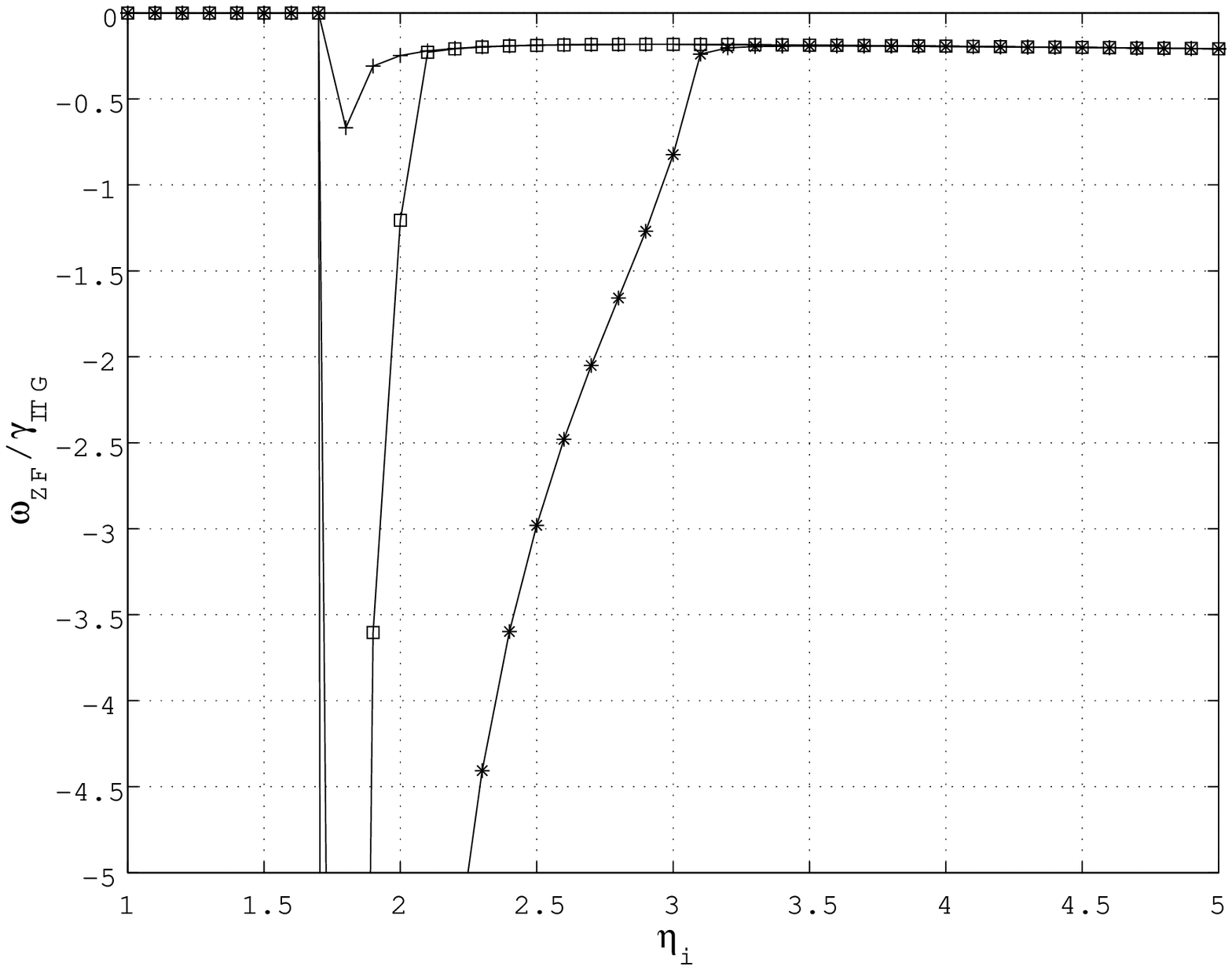}
  \caption{(2 and 3) The zonal flow growth rate (Figure 2) and frequency (Figure 3) as a function of $\eta_i$ with $q_f |\Phi_f|$ as a parameter. The other parameters are as $q_x = 0.3 = k_x = k_y$, $\epsilon_n = 1.0$ and $\tau = 1$. The results are shown for $q_f |\Phi_f| = 0$ (plus), $q_f |\Phi_f| = 100$ (boxes) and $q_f |\Phi_f| = 500$ (asterisk).}
\end{figure}

In Figure 4 and 5 the zonal flow growth rate (Figure 4) and real frequency (Figure 5) (normalized to the linear ITG growth rate) as a function of $\epsilon_n$ is shown with mean flow shear as a parameter is displayed. The other parameters are as in Figure 1 with $\eta_i = 4$ and $q_f = 0.1$. The results are shown for $|\Phi_f| = 0$ (asterisk), $|\Phi_f| = 100$ (diamonds) and $|\Phi_f| = 500$ (stars). Note, that in the present normalization the expansion term ($k_y q_f^2 |\Phi_f||$) is still small compared to the growth rate in the weak flow case while it is comparable to the growth rate in the strong flow case. The results show a suppression of zonal flow growth with $|\Phi_f|$. The reason for this is in the variation of the radial group velocity ($v_{gx}$) in the total time frame,  as in the decorrelation of drift wave propagation by a sheared flow. This will weaken the modulation response of the drift wave response. The stabilization of zonal flow growth by the sheared flow results in a significant increase of the real frequency ($|\Omega_r|$). The effects resulting from from this approach is in qualitative agreement with with results reported earlier using other drift wave models~\cite{a48} and using the coherent mode coupling method in ETG background turbulence for generating the zonal flow~\cite{a61}.

\begin{figure}
  \includegraphics[height=.3\textheight]{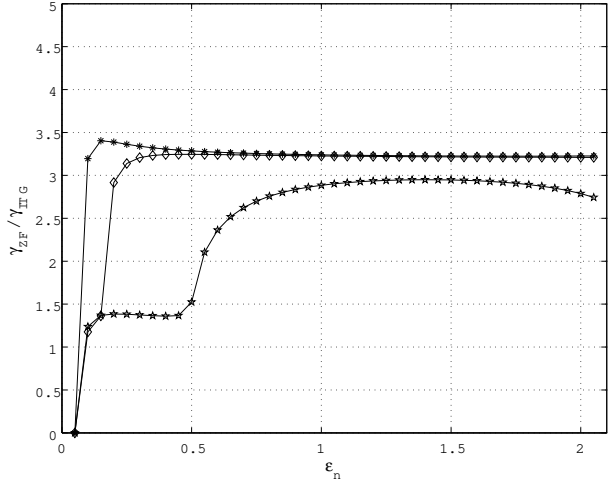}
  \caption{(4 and 5) The zonal flow growth rate (Figure 4) and frequency (Figure 5) (normalized to the linear ITG growth rate) as a function of $\epsilon_n$ is shown with mean flow shear as a parameter. The other parameters are $q_x = 0.3 = k_x = k_y$, $\tau = 1$ and $\eta_i = 3$. The results are displayed for $q_f |\Phi_f| = 0$ (asterisk), $q_f |\Phi_f| = 100$ (diamonds) and $q_f |\Phi_f| = 500$ (stars).}
\end{figure}

\begin{figure}
  \includegraphics[height=.3\textheight]{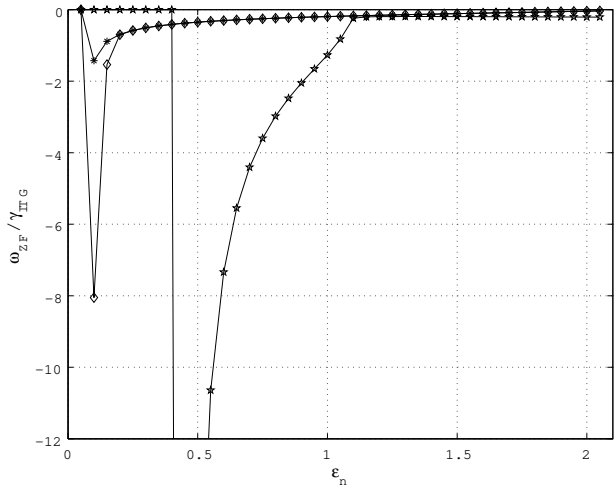}
  \caption{(4 and 5) The zonal flow growth rate (Figure 4) and frequency (Figure 5) (normalized to the linear ITG growth rate) as a function of $\epsilon_n$ is shown with mean flow shear as a parameter. The other parameters are $q_x = 0.3 = k_x = k_y$, $\tau = 1$ and $\eta_i = 3$. The results are displayed for $q_f |\Phi_f| = 0$ (asterisk), $q_f |\Phi_f| = 100$ (diamonds) and $q_f |\Phi_f| = 500$ (stars).}
\end{figure}

Next, the results using the extended model including parallel ion momentum is treated (solutions to the dispersion relation Eq. 28). 

In Figure 6, the ratio of the growth rate and real frequency is shown as a function of $\eta_i$ with safety factor ($q$) as a parameter. In the present case the safety factor is varied from $q=2$ (asterisk curve) to $q=8$ (plus curve). The other parameters are as in Figure 1. It is found that the ratio of the growth rate and the frequency is decreasing with increasing $\eta_i$ (remember that $L_n$ is considered to be fixed). This is suggestive of a transition from a state of stable zonal flows to a state with oscillating zonal flows. It is indicated that in the region close to the linear ITG threshold the zonal flows are stationary and may have a significant stabilizing effect on the background turbulence whereas at higher $\eta_i$ the zonal flows becomes oscillatory. In general, in a comparison of the previous model and the present model the change in zonal flow generation inherently comes from the fact that $|\omega_r|$ is increased in the system including parallel ion momentum, whereas the ITG growth rates are rather similar in both systems. 

\begin{figure}
  \includegraphics[height=.3\textheight]{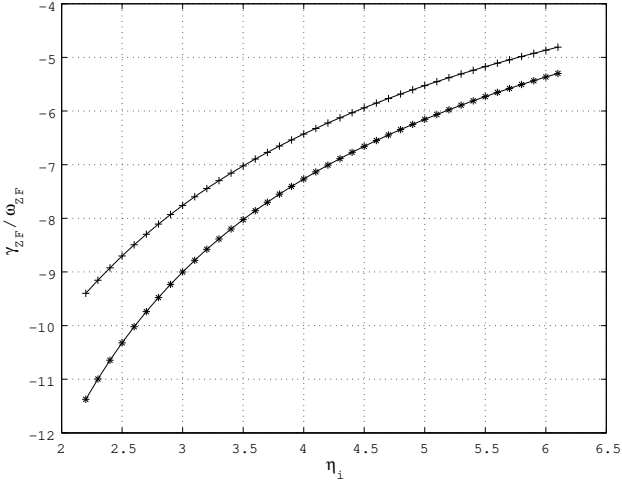}
  \caption{The ratio of the growth rate and frequency $\gamma_{ZF}/\omega_{ZF}$ as a function of $\eta_i$ with safety factor ($q$) as a parameter. The other parameters are as $q_x = 0.3 = k_x = k_y$, $\epsilon_n = 1.0$, $\tau = 1$ and $q_f |\Phi_f| = 0$. The results are displayed for $q=2$ (asterisk curve) to $q=8$ (plus curve).}
\end{figure}

\section{Summary}
This work focus mainly on analytical estimations of zonal flow growth rate and real frequency under the influence of sheared mean flows and parallel ion motion. An analytical model for the generation of zonal flows by ion-temperature-gradient background turbulence in the presence of a sheared mean flow and parallel ion motion are derived using the wave kinetic approach. The model consists of the ion continuity and ion temperature equations, in addition, the parallel momentum equation for the ions are included. The zonal flow evolution is described by the vorticity equation including a collisional damping on the zonal flow generation. The zonal flow growth rates and real frequencies are scanned for a wide range of plasma parameters.

It was found the general level of zonal flow was suppressed by a sheared background flow and also that for obtaining realistic results retaining collisions into the analytical model seems to be very important. 

The results show a suppression of zonal flow growth with $|\Phi_f|$. This is in agreement with previous research of zonal flow generation using another drift wave model~\cite{a48} and the coherent mode coupling method in ETG drift wave turbulence~\cite{a61}. The reason for this is in the variation of the radial group velocity ($v_{gx}$) in the total time frame,  as in the decorrelation of drift wave propagation by a sheared flow. This will weaken the modulation response of the drift wave response. The stabilization of zonal flow growth by the sheared flow results in a significant increase of the real frequency ($|\Omega_r|$). 

By introducing collisional damping a suppression of the zonal flow growth rate is found.  In the case of no damping $\mu = 0$, no explicit $\eta_i$ dependency on the zonal flow growth rate and real frequency appears, whereas, a modest effect of damping is exhibited. 

In addition, it is found that the parallel ion motion may reduce the ZF generation significantly. This is suggestive of a transition from a state of stable zonal flows to a state with oscillating zonal flows. It is indicated that in the region close to the linear ITG threshold the zonal flows are stable and may have a significant stabilizing effect on the background turbulence whereas at higher $\eta_i$ the zonal flows becomes oscillatory. In general, in a comparison of the previous model and the present model the change in zonal flow generation inherently comes from the fact that $|\omega_r|$ is increased in the system including parallel ion momentum, whereas the ITG growth rates are rather similar in both systems. 
\section{Acknowledgment}
This research was supported by Japan Society for the Promotion of Science (JSPS). The sponsors do not bear any responsibility for the contents in this work. The Authors are indebted to Professor J. Li for fruitful discussions.

\end{document}